\documentstyle[preprint,aps]{revtex}
\begin{document}
\draft
\title{Factorization Contributions and the Breaking of the $\Delta I=1/2$\\
Rule in Weak
$\Lambda N\rho$ and $\Sigma N\rho$ Couplings \\
\begin{flushright}
ADP-95-5/T172 \\
nucl-th/9504006
\end{flushright}}
\author{Kim Maltman\cite{byline}}
\address{Department of Mathematics and Statistics, York University,
4700 Keele St., \\ North York, Ontario, Canada M3J 1P3}
\author{Mikhail Shmatikov}
\address{Russian Research Center ``Kurchatov Institute'', 123182, Moscow,
Russia}
\date{\today}
\maketitle
\begin{abstract}
We compute the modified factorization contributions to the $\Lambda\rightarrow
N\rho$ and $\Sigma\rightarrow N\rho$ couplings and demonstrate
that these contributions naturally include $\Delta I=3/2$ terms which
are comparable ($\simeq 0.4$ to $-0.8$ times)
in magnitude to the
corresponding $\Delta I=1/2$ terms.  As a consequence, we conclude
that models which treat vector meson exchange contributions to the
weak conversion process $\Lambda N\rightarrow NN$ assuming
such weak couplings to satisfy the $\Delta I=1/2$ rule are unlikely
to be reliable.
\end{abstract}

The $\Delta I=1/2\ $ rule is a prominent feature of observed
$\Delta S=1\ $ non-leptonic weak interactions ($K$ decay and hyperon decay).
Not only is the ratio of $\Delta I=1/2\ $ to $\Delta I=3/2\ $
amplitudes considerably enhanced over that of the corresponding
un-QCD-modified operator strengths, but also the non-leptonic decays
completely dominate semi-leptonic decay modes, indicating
a significant enhancement of the $\Delta I=1/2\ $ amplitudes.  As a
consequence of this observation, it has become conventional, in the
absence of other evidence, to assume the validity of the $\Delta I=1/2\ $ rule
for all $\Delta S=1\ $ non-leptonic weak interactions.  In
particular, in the meson-exchange treatment of $\Lambda N\rightarrow NN\ $
it has been assumed that the relevant weak
baryon-meson couplings satisfy the rule.  In the case of the
$\pi$ couplings, this is known empirically, from hyperon decay,
to be a valid assumption, but no similar experimental
support exists for the assumption that vector meson couplings satisfy
the rule.  In this note we argue that, for the latter couplings, one
may indeed expect significant violations of the $\Delta I=1/2\ $ rule.
We base this statement on an evaluation of factorization
contributions to the couplings and show below how, for such
contributions, the structure of QCD modifications to the weak
interactions are such as to naturally distinguish the
pseudoscalar and vector cases.

As is well-known, the effects of QCD on the $\Delta S=1\ $
non-leptonic interactions can be taken into account perturbatively,
down to a scale $\simeq 1$ GeV where the strong interactions
begin to become truly strong \cite{ref1,ref2,ref3,ref4} .
One obtains, for the effective
$\Delta S=1\ $ non-leptonic Hamiltonian
\begin{equation}
{\cal H}_{eff}=-\sqrt{2}G\sin\theta_C\cos\theta_C
\sum^{6}_{i=1}c_i O_i
\label{one}
\end{equation}
where the operators, $O_i$, have the form
\begin{eqnarray}
&&O_1 = \bar{d}_L\gamma_{\mu}s_L\,\bar{u}_L\gamma^{\mu}u_L\, -\,
\bar{u}_L\gamma_{\mu}s_L\,\bar{d}_L\gamma^{\mu}u_L
\nonumber \\
&&O_2 = \bar{d}_L\gamma_{\mu}s_L\,\bar{u}_L\gamma^{\mu}u_L\, +\,
\bar{u}_L\gamma_{\mu}s_L\,\bar{d}_L\gamma^{\mu}u_L\,+\,
2\,\bar{d}_L\gamma_{\mu}s_L\,\bar{d}_L\gamma^{\mu}d_L+\,
2\,\bar{d}_L\gamma_{\mu}s_L\,\bar{s}_L\gamma^{\mu}s_L\nonumber \\
&&O_3 = \bar{d}_L\gamma_{\mu}s_L\,\bar{u}_L\gamma^{\mu}u_L\, +\,
\bar{u}_L\gamma_{\mu}s_L\,\bar{d}_L\gamma^{\mu}u_L\,+\,
2\,\bar{d}_L\gamma_{\mu}s_L\,\bar{d}_L\gamma^{\mu}d_L-\,
3\,\bar{d}_L\gamma_{\mu}s_L\,\bar{s}_L\gamma^{\mu}s_L\nonumber \\
&&O_4 = \bar{d}_L\gamma_{\mu}s_L\,\bar{u}_L\gamma^{\mu}u_L\, +\,
\bar{u}_L\gamma_{\mu}s_L\,\bar{d}_L\gamma^{\mu}u_L\,-\,
\bar{d}_L\gamma_{\mu}s_L\,\bar{d}_L\gamma^{\mu}d_L\nonumber \\
&&O_5 = \bar{d}_L\gamma_{\mu}\lambda^a s_L(\bar{u}_R\gamma^{\mu}
\lambda^a u_R\, +\, \bar{d}_R\gamma^{\mu}\lambda^a d_R\, +\,
\bar{s}_R\gamma^{\mu}\lambda^a s_R)\nonumber \\
&&O_6 = \bar{d}_L\gamma_{\mu}s_L(\bar{u}_R\gamma^{\mu}u_R\, +\,
\bar{d}_R\gamma^{\mu}d_R\, +\, \bar{s}_R\gamma^{\mu}s_R)
\label{two}
\end{eqnarray}
and the coefficients, $c_i$, are scale-dependent and calculable
perturbatively.  The operators $O_1,\cdots ,O_6$ in Eq. (\ref{two})
have the specific $(flavor, isospin)$ quantum numbers
$(8, 1/2)$, $(8, 1/2)$, $(27, 1/2)$, $(27, 3/2)$, $(8, 1/2)$ and
$(8, 1/2)$, respectively.  The operators $O_{5,6}$, with LR
chiral structure are due to penguin graphs.  The leading ($O_1$) term
typically has a coefficient, $c_1\simeq 4\, c_4$ at a scale $\simeq 1$ GeV,
$O_4$ being the only $\Delta I=3/2\ $ operator,
indicating that a portion of the observed experimental enhancement
results from QCD modifications of the relative operator strengths.
The additional factor of $\simeq 4-5$ in the
observed amplitude ratios not accounted for by this modification,
however, must be associated with specific dynamics in
the matrix elements of the operators.  In the case of $K$ decay, it
seems likely that a significant portion of this dynamical enhancement
is associated with final state interactions (FSI), the
$\Delta I=1/2\ $ operators leading to the attractive
$I=0$ $\pi\pi$ s-wave final state, the $\Delta I=3/2\ $
operator to the replusive $I=2$ state \cite{ref5,ref6,ref7}.
A similar explanation is not, however, tenable for hyperon decays, since,
at least for $\Lambda\ $ and $\Sigma\ $, the final
state phases are known to be small.  An old idea \cite{ref8}
which provides an attractive (if
qualitative) alternative
for these decays, is based on the observation that there are
large enhancements of the penguin operator matrix elements in
what is usually called the factorization approximation, these
enhancements resulting from the different, LR, chiral structure
of these operators.  We briefly describe this approximation below.

In the approximation that FSI may be neglected, one may separate
the graphs contributing to baryon-meson, $B^\prime\rightarrow BM$,
weak transitions into two classes: ``external'', describing those
graphs in which both a quark and anti-quark line from the effective
quark-level weak vertex end up in the final state meson, $M$, and
``internal'', describing all other graphs.  The advantage of this
classification is that the external contributions are effectively
``factorized'' into a product of the matrix elements of two currents,
one connecting $B^\prime$ to $B$, and one connecting $M$
to the vacuum.  These current matrix elements are completely known
in terms of baryon semileptonic decay form factors and meson
decay constants, so that the ``external'', or ``factorization''
contributions may be reliably calculated.  This is not
the case for the internal contributions.  (Note that it is
essential to use the modified form of factorization, in which
Fierz rearrangements of the $O_i$ are also taken into account,
or one will fail to satisfy the correct isospin relations between
factorization matrix elements of the operators.)  As mentioned
above, when $M$ is a pion, the Fierz-rearranged contributions
of $O_{5,6}$ (the non-rearranged
expectation is zero because the $\Delta S=1$ current portion of
the non-rearranged form is a flavor singlet) contain a large
enhancement relative to the matrix elements of the LL operators.
Since these operators are pure $\Delta I=1/2\ $
this provides an attractive qualitative explanation of the
$\Delta I=1/2\ $ rule, especially when combined with the observation
that the $\Delta I=3/2\ $ pieces of the internal contributions would
vanish in the naive quark model limit (in which the baryons contain
only the leading three-quark color-singlet Fock space
component) owing to the color symmetry of the $\Delta I=3/2\ $
operator, $O_4$.  The prescription of simply ignoring the
internal contributions is called the ``factorization
approximation''.  The predictions of this approximation
are actually rather ill-defined, since the values of the
Wilson coefficients of the penguin operators, which
arise from the evolution below a scale $\simeq m_c$, are quite
sensitive to the precise scale chosen, making
amplitudes where these terms occur multiplied by a
large enhancement factor also quite sensitive to the
scale choice.  What can more safely be determined are
factorization contributions to quantities which do not involve
the (enhanced) penguin operators, for example, the $\Delta I=3/2\ $
contributions to hyperon decay.  Here, if one takes the
coefficient $c_4$ to be evaluated at a scale of $1$ GeV and uses
the true $\Delta I=3/2\ $
amplitudes obtained after making corrections for $\Sigma$-$\Lambda$
and $\pi_3$-$\pi_8$ mixing in the physical amplitudes \cite{ref9} (the
p-wave $\Lambda$ and $\Xi$ amplitudes are increased by $\simeq 400\%$
and $\simeq 100\%$ by these corrections),
one
finds (1) good fits to the s- and p-wave $\Delta I=3/2\ $
$\Xi$ amplitudes and s-wave $\Sigma\ $
triangle discrepancy
(2) that the p-wave $\Sigma\ $ triangle
discrepancy is underestimated by a factor of $4$, and (3) that
the s- and p-wave $\Lambda\ $ amplitudes
are overestimated by a factor of $3-4$ (the p-wave factorization
contribution also being opposite in sign to the experimental value).
Although one should bear in mind that the experimental
errors on the $\Delta I=3/2\ $ amplitudes
are rather large (apart from the $\Lambda\ $ s-wave,
the factorization predictions fall within $\simeq 2\sigma$ of the
central experimental value), it seems safe to conclude that, while
the factorization contributions are of roughly the correct magnitude,
there are additional non-negligible contributions from the
internal graphs (even for $\Delta I=3/2$).
In what follows we will be treating weak vector meson couplings
for which, as we will see below, the factorization contributions
of the penguin operators vanish.  We will then see that the
remaining factorization contributions involve large violations
of the $\Delta I=1/2\ $ rule and,
in light of the above discussion, argue that one should, therefore,
expect some portion of this violation to survive in the
total couplings.

Let us turn to the evaluation of the factorization contributions
to the weak $\Lambda N\rho\ $ and $\Sigma N\rho\ $
couplings (the corresponding contributions to both $\Lambda N\omega$
and $\Sigma N\omega$ couplings are small and, even for
$\Sigma N\omega$, satisfy the $\Delta I=1/2\ $
rule, so we will not discuss them further).  We employ the effective
weak Hamiltonian of Eqs. (1), (2), with coefficients, $c_i$,
evaluated at a scale $1$ GeV \cite{ref4,ref10} :  $c_1=-1.90$, $c_2=0.14$,
$c_3=0.10$, $c_4=0.49$.  The coefficients $c_{5,6}$ are not needed since
the factorization contributions to the weak $\rho$ couplings of the operators
$O_{5,6}$ vanish.  This fact follows from the observation that
these operators contain color-singlet flavor-octet non-strange
currents only in Fierz-rearranged form, in which form, owing to the
original LR chiral structure, only scalar
and pseudoscalar currents are involved,
the vacuum-to-$\rho$ matrix elements of which
automatically vanish.  The remaining contributions are straightforward
to work out.

We define the effective weak couplings via
\begin{equation}
<N(p^\prime )\rho\vert H_{eff}\vert Y(p)> =\epsilon^{(\rho )*}_\mu
\bar u_N(p^\prime )\biggl[
f_1^w\gamma^\mu -i{\sigma^{\mu\nu}q_\nu\over 2m_N}f_2^w
+g_1^w\gamma^\mu\gamma_5 -i{\sigma^{\mu\nu}q_\nu\over 2m_N}g_2^w
\gamma_5\biggr]
u_Y(p)\label{three}
\end{equation}
where $q=p-p^\prime $ and $\epsilon^{(\rho )}_\mu$ is the $\rho$
polarization vector, and the baryon transition form factors via
\begin{equation}
<B^\prime (p^\prime )\vert\  V_\mu -A_\mu\ \vert B(p)>=
\bar{u}_{B^\prime}(p^\prime )\biggl[ f_1\gamma_\mu -i{\sigma_{\mu\nu}q^\nu
\over 2m_N}f_2 +g_1\gamma_\mu\gamma_5 +i{q_\mu\gamma_5\over 2m_N}g_3
\biggr] u_B(p)\ \label{four}
\end{equation}
where we have dropped the $2^{nd}$ class current form factors, $f_3$
and $g_2$ in Eq. (4).  In using Eq. (4) below, we will assume
that $f_1$ and $f_2$ are given by their CVC values, and take
$g_1/f_1$ from hyperon semi-leptonic decay data \cite{ref11} .
$g_3$ does not enter the expressions for the factorization
contributions to the couplings due to the transversality
of the $\rho$ polarization vector.  Defining the $\rho$ decay
constant, $f_\rho$, by
\begin{equation}
<O\vert V^3_\mu\vert \rho^o(\vec{q})>=f_\rho m^2_\rho\epsilon^{(\rho )}
_\mu (\vec{q}),
\label{five}
\end{equation}
we then obtain for the factorization contributions to the
$\Lambda\rightarrow p\rho^-$ couplings
\begin{eqnarray}
&&f_i^w=\sqrt{2}K\, [-{2\over 3}c_1+{4\over 3}c_2+{4\over 3}c_3
+{4\over 3}c_4]\, f_i^{\Lambda p}\quad (i=1,2)\nonumber \\
&&g_1^w=f_1^w\, [g_1^{\Lambda p}/f_1^{\Lambda p}]\label{six}
\end{eqnarray}
with
\begin{equation}
K={G_F\over 4\sqrt{2}}\sin (2\theta_c)f_\rho m^2_\rho\ ,\label{seven}
\end{equation}
where $G_F$, $\theta_c$ are the Fermi constant and Cabibbo angle,
respectively, and $f_i^{\Lambda p}$, $g_i^{\Lambda p}$ the
form factors relevant to
$<p\vert\bar{u}(\gamma_\mu -\gamma_\mu\gamma_5)s\vert \Lambda >$.
We have, from CVC, $f_1^{\Lambda p}(0)=-\sqrt{3\over 2}$,
$f_2^{\Lambda p}(0)/f_1^{\Lambda p}(0)=1.63$ and, from $\Lambda\ $
semileptonic decay data, $g_1^{\Lambda p}(0)/f_1^{\Lambda p}(0)=-0.72$.
As mentioned above, $g_2^{B^\prime B}$ is $2^{nd}$ class, and assumed
to be zero.
Since $O_1, \cdots ,O_3$ are $\Delta I=1/2\ $
and $O_4$ $\Delta I=3/2\ $
the corresponding $\Lambda\rightarrow n\rho^o\ $ contributions
follow from isospin Clebsch-Gordan coefficients,
leading to
\begin{eqnarray}
&&f_i^w=-K\, [-{2\over 3}c_1+{4\over 3}c_2+{4\over 3}c_3
-{8\over 3}c_4]\, f_i^{\Lambda p}\quad (i=1,2)\nonumber \\
&&g_1^w=f_1^w\, [g_1^{\Lambda p}/f_1^{\Lambda p}]\ .\label{eight}
\end{eqnarray}
One finds similarly, for the factorization contributions to the
$\Sigma^-\rightarrow n\rho^-\ $ couplings,
\begin{eqnarray}
&&f_i^w=\sqrt{2}K\, [-{2\over 3}c_1+{4\over 3}c_2+{4\over 3}c_3
+{4\over 3}c_4]\,  f_i^{\Sigma^-n}\quad (i=1,2)\nonumber \\
&&g_1^w=f_1^w\, [g_1^{\Sigma^-n}/f_1^{\Sigma^-n}]\ ,\label{nine}
\end{eqnarray}
where, from CVC, $f_1^{\Sigma^-n}(0)=-1$,
$f_2^{\Sigma^-n}(0)/f_1^{\Sigma^-n}(0)=-1.86$ (compatible with
experiment \cite{ref11} )
and, from experiment,
$g_1^{\Sigma^-n}(0)/f_1^{\Sigma^-n}(0)=0.34$ \cite{ref11} .  From the form
of the operators $O_1,\cdots ,O_4$, one sees that there are no
terms containing simultaneously both a $\bar{u}$ and $d$ field, and
hence both the $\Delta I=1/2\ $ and $\Delta I=3/2\ $
factorization contributions to the $\Sigma^+\rightarrow n\rho^+\ $
couplings vanish.  This reduces the number of independent reduced
matrix elements for both the $\Delta I=1/2\ $ and $\Delta I=3/2$ operators
from two to one, and one may then show that
the $\Delta I=1/2\ $
factorization contributions to the $\Sigma N\rho\ $
couplings are in the ratios $1:{1\over \sqrt{2}}:{1\over\sqrt{2}}:{1\over 2}$
and the $\Delta I=3/2\ $
contributions in the ratios
$1:-\sqrt{2}:{1\over\sqrt{2}}:-1$,
for $\Sigma^-\rightarrow n\rho^-\ $ ,
$\Sigma^+\rightarrow p\rho^o\ $ , $\Sigma^o\rightarrow p\rho^-\ $ ,
and $\Sigma^o\rightarrow n\rho^o\ $ , respectively.
Expressions for the weak $\Sigma^+\rightarrow p\rho^o\ $ ,
$\Sigma^o\rightarrow p\rho^-\ $
and $\Sigma^o\rightarrow n\rho^o\ $ couplings are then
readily obtained from those of Eq. (9).

{}From the expressions (6), (8), (9), and the discussion
below Eq. (9), we see that the relative strength of the
$\Delta I=1/2\ $ to $\Delta I=3/2\ $
contributions to the weak couplings is determined, in all cases,
by the factor $[-{2\over 3}c_1+{4\over 3}c_2+{4\over 3}c_3
+{4\over 3}c_4] $ for the $\rho^-$ couplings, and
$[-{2\over 3}c_1+{4\over 3}c_2+{4\over 3}c_3
-{8\over 3}c_4] $ for the $\rho^o$ couplings.
For the $\rho^-$ couplings, using the $c_i$ values quoted above
(corresponding to a scale $1$ GeV), the contributions of
the $27_F$ $c_3$ and $c_4$ terms are $0.54$ times those
of the $8_F$ $c_1$ and $c_2$ terms.  The leading $\Delta I=3/2\ $
term is $0.52$ times the leading $\Delta I=1/2\ $
term and $0.41$ times the net $\Delta I=1/2\ $
contribution.  Similarly, for the $\rho^o$ couplings,
the $27_F$ contributions are $-0.81$ times the $8_F$ contributions,
while the leading $\Delta I=3/2\ $
contribution is $-1.03$ times the leading $\Delta I=1/2\ $
contribution and $-0.82$ times the net $\Delta I=1/2\ $
contribution.  The factorization contributions to the weak
$\Lambda N\rho\ $ and $\Sigma N\rho\ $
couplings thus badly violate the $\Delta I=1/2\ $
rule.  The basic reason for this is the complete absence
of the penguin contributions,
which had large enhancements in the $\pi$ coupling case.

We present the numerical results for the factorization contributions
to the $\Lambda N\rho\ $
and $\Sigma N\rho\ $
weak couplings $f_1^w$, $f_2^w$ and $g_1^w$ in Table I.  The ratios
reflect the strong violation of the $\Delta I=1/2\ $
rule discussed above.  This violation will, however, be reflected in
the full couplings only if the factorization contributions
represent a moderate to sizable fraction of the full couplings.
To see whether or not this is likely to be the case, we
consider two existing models which have made predictions for the
$\Lambda N\rho\ $
(though not the $\Sigma N\rho$)
couplings \cite{ref12,ref13} .  In Table II we compare
the factorization contributions to the
$\Lambda\rightarrow p\rho^-\ $ couplings
obtained above with the values obtained in the models of
Refs.\ \onlinecite{ref12,ref13} (adjusted to Particle Data
Group conventions for $\gamma_5$).  We will return to a brief discussion
of the models below, but for the moment, two features of the
table are of note.  First, the model predictions differ
considerably, most significantly for the parity-violating
$g_1^w$ coupling.  Second, the factorization contributions
are $\simeq 1/3$ of the full predictions for $f_1^w$, $f_2^w$
and between $\simeq 1/4$ and $1$ times the full prediction
for $g_1^w$.  Also, as we will discuss below, there are significant
uncertainties in the model predictions.  From Table II it thus
appears to us extremely unlikely that one can ignore the
$\Delta I=3/2\ $
components of the
$\Lambda N\rho\ $ and $\Sigma N\rho\ $
couplings.

A few words are in order concerning the models of
Refs.\ \onlinecite{ref12,ref13} which we have used to
gauge the potential importance of the factorization
contributions.  The model of Ref.\ \onlinecite{ref12}
provides the framework for the
weak couplings of the meson-exchange treatment of
$\Lambda N\rightarrow NN\ $ employed
by Dubach et al. \cite{ref14}.  Here the parity violating (PV)
$g_1^w$ coupling is obtained from the known PV $\Lambda p\pi^-$
and $\Sigma^+p\pi^o$ couplings via the $SU(6)_w$ treatment
of Desplanques, Donoghue and Holstein \cite{ref15} , where
factorization estimates have been used to provide values
for the two $SU(6)_w$ reduced matrix elements,
$a_V$ and $a_T$, present in the $\Lambda N\rho\ $
couplings but not in the $\Lambda N\pi$ and $\Sigma N\pi$
couplings \cite{ref15} .  The parity conserving (PC)
$f_1^w$, $f_2^w$ couplings are obtained via a pole model
analysis which includes ground state baryon pole terms and
$K^*$ pole terms, the strong vector meson couplings
being obtained from an $SU(3)_F$ vector dominance model (VDM)
treatment together with the weak baryon-baryon transition
matrix elements from an analogous treatment of the PC
$\Lambda \rightarrow N\pi$ and $\Sigma\rightarrow N\pi$
amplitudes.  In Ref.\ \onlinecite{ref13} , the PV $g_1^w$
coupling is obtained from a pole model treatment which
keeps only baryon poles belonging to the $(70,1^-)$ multiplet
of (ordinary) $SU(6)$.  The required weak baryon transition
matrix elements between ground state and negative
parity excited state baryons are taken from a
treatment \cite{ref16} of hyperon s-wave $\pi$ decays
which includes the leading commutator terms as well as
the negative parity baryon poles, and which fixes the
PV baryon-baryon matrix elements by assuming (1) that the $D/F$
ratio for the weak baryon transitions is $-1$ and (2) that
experimentally observed deviations from a modified Lee-Sugawara
sum rule are due entirely to the negative parity baryon poles.
The required strong couplings are obtained using VDM arguments,
together with information on the scalar multipoles in
$S_{11}$, $S_{11}^\prime$ electroproduction.  The PC $f_1^w$,
$f_2^w$ couplings are obtained using a pole model treatment
which includes ${1\over 2}^{+*}$ baryon resonance pole
contributions, in addition to the ground state baryon pole and $K^*$
pole terms of Ref.\ \onlinecite{ref12} .  The strong couplings
required are obtained again using VDM arguments, together
with data on $P_{11}$ radiative decays and assumptions about
the scalar multipoles in $P_{11}$ electroproduction.  The
weak baryon-baryon couplings for the ${1\over 2}^{+*}$ pole
terms are taken from a pole term analysis of hyperon p-wave
$\pi$ decays\ \onlinecite{ref16} which (1) assumes an $F/D$
ratio of $-1$ for the weak baryon transitions and (2) fixes
the overall strength by optimizing the full fit to the
experimental p-wave amplitudes.  This fit, however, employs
a $K-\pi$ weak transition strength in its $K$-pole graphs
an order of magnitude greater than that extracted from $K\rightarrow \pi\pi$
\cite{ref17} , which makes the whole procedure appear somewhat
dubious.  The ${1\over 2}^{+*}$ pole contributions to
$f_1^w$ in Ref.\ \onlinecite{ref13} are negligible, but this is
not true of the corresponding contributions to $f_2^w$.
Finally, the $K^*$ pole contributions are obtained using VDM plus
$SU(3)_F$ arguments for the strong $K^*$ couplings and a factorization
treatment, which keeps only the $O_1$, $O_2$ terms of ${\cal{H}}_{eff}$
and drops the Fierz-rearranged contributions, for the $K^*-\rho$
weak transition.  This is, in fact, a rather suspect way to
treat factorization contributions, even if they were expected
to represent well the full coupling.  Indeed, if one keeps all
terms in ${\cal{H}}_{eff}$, one finds the same linear combinations
of the $c_i$ occuring for the charged and neutral $K^*-\rho$
mixing terms as occur for the charged and neutral $\rho$
weak couplings above; i.e., there is very significant breaking
of the $\Delta I=1/2\ $
rule for the $K^*$ pole terms.

As can be seen from the discussion above, there are many assumptions
and approximations which enter the models of Refs.\ \onlinecite{ref12,ref13} .
As such, the model values for the weak couplings quoted in Table II
will involve significant uncertainties whose sizes are difficult to
quantify.  We feel, however, that the arguments leading to the
ground state baryon pole contributions, especially to $f_1^w$,
are likely to be the most reliable, so these contributions provide
a useful benchmark.  For $\Lambda\rightarrow p\rho^-$, these are
$-10.5\times 10^{-7}$ and $-11.3\times 10^{-7}$ for the models
of Refs.\ \onlinecite{ref12,ref13} , respectively.
Similarly, for the $f_2^w$ $\Lambda\rightarrow p\rho^-$ coupling,
these contributions are $-14.8\times 10^{-7}$ and $-9.3\times 10^{-7}$,
respectively.  Note that only the $n$ pole term contributes to
$f_1^w$, but that both $n$ and $\Sigma^+$ pole terms contribute
to $f_2^w$.  In the latter case, there is considerable
cancellation between the two terms, which makes the actual
result rather sensitive to possible $SU(3)_F$ breaking in the
relative strengths of the strong $pn\rho^-$ and $\Sigma^+\Lambda\rho^-$
couplings, a $\pm 30\%$ variation of the the $\Sigma^+\Lambda\rho^-$
strength from its $SU(3)_F$ value, for example, producing a variation
of $\pm 7\times 10^{-7}$ ($\pm 10\times 10^{-7}$)
in the corresponding ground state
baryon pole contribution to $f_2^w$
for the parametrizations of Refs.\ \onlinecite{ref12,ref13} ,
respectively.  One should also note that the
close agreement of the $f_2^w$ values in the two models is actually
a numerical accident, since the ${1\over 2}^{+*}$ baryon pole terms
of Ref.\ \onlinecite{ref13} , completely absent in Ref.\ \onlinecite{ref12} ,
contribute $\simeq 1/3$ of the quoted $f_2^w$ value.

To summarize, we have evaluated the factorization contributions
to the weak $\Lambda N\rho\ $
and $\Sigma N\rho\ $
couplings and find that they involve large
violations of the $\Delta I=1/2\ $
rule.  Since the size of these contributions is not small on
the scale of values to be expected for the full couplings, it
appears very unlikely that it is safe to assume the validity of the
$\Delta I=1/2\ $ rule as an input when determining the
weak $\Lambda N\rho\ $
couplings to be used in treating $\Lambda N\rightarrow NN\ $
in the meson-exchange framework.

\acknowledgements
\noindent
KM would like to acknowledge the support of the Natural Sciences
and Engineering Research Council of Canada and thank the Department
of Physics and Mathematical Physics of the University of Adelaide
for its hospitality during the course of this work.  Both authors
acknowledge Angels Ramos for useful discussion concerning the
content of Ref.\ \onlinecite{ref12}.

\begin{table}
\caption{Factorization contributions to the weak $\Lambda N\rho$,
$\Sigma N\rho$ couplings.  All entries in units of $10^{-7}$.
\label{Table1}}
\begin{tabular}{lrrr}
Process&$f_1^w$&$f_2^w$&$g_1^w$ \\
\tableline
$\Lambda\rightarrow p\rho^-$&-4.1&-6.6&2.9 \\
$\Lambda\rightarrow n\rho^o$&0.36&0.58&-0.26 \\
$\Sigma^+\rightarrow p\rho^o$&-0.29&0.54&-0.10 \\
$\Sigma^+\rightarrow n\rho^+$&0&0&0 \\
$\Sigma^o\rightarrow p\rho^-$&-2.3&4.3&-0.79 \\
$\Sigma^o\rightarrow n\rho^o$&-0.21&0.38&-0.07 \\
$\Sigma^-\rightarrow n\rho^-$&-3.3&6.1&-1.1 \\
\end{tabular}
\end{table}

\begin{table}
\caption{Comparison of $q^2=0$ factorization and model values for
$\Lambda p\rho^-$
couplings.  All entries in units of $10^{-7}$.  The models
are as discussed in the text.
\label{Table2}}
\begin{tabular}{lccc}
Coupling&Factorization&Ref.\ \onlinecite{ref12}&Ref.\ \onlinecite{ref13} \\
\tableline
$f_1^w$&-4.1&-15.0&-10.6 \\
$f_2^w$&-6.6&-22.6&-24.3 \\
$g_1^w$&2.9&3.4&12.0 \\
\end{tabular}
\end{table}
\end{document}